\documentclass[conference,a4paper]{IEEEtran}
\IEEEoverridecommandlockouts
\usepackage{cite}
\usepackage{amsmath,amssymb,amsfonts}
\usepackage{algorithmic}
\usepackage{graphicx}
\usepackage{textcomp}
\usepackage{xcolor}
\usepackage[hyphens]{url}
\usepackage{comment}
\def\BibTeX{{\rm B\kern-.05em{\sc i\kern-.025em b}\kern-.08em
    T\kern-.1667em\lower.7ex\hbox{E}\kern-.125emX}}
\begin{document}
\title{Cervical Cancer Detection Using Multi-Branch Deep Learning Model\\}

\author{
    \IEEEauthorblockN{1\textsuperscript{st} Tatsuhiro Baba}
    \IEEEauthorblockA{
        \textit{School of Computer Science and Engineering} \\
        \textit{University of Aizu}, Fukushima, Japan \\
        s1290244@u-aizu.ac.jp 
    }
    \and
    \IEEEauthorblockN{2\textsuperscript{nd} Abu Saleh Musa Miah}
    \IEEEauthorblockA{
        \textit{School of Computer Science and Engineering} \\
        \textit{University of Aizu}, Fukushima, Japan \\
        musa@u-aizu.ac.jp
    }
    \and
    \IEEEauthorblockN{3\textsuperscript{rd} Jungpil Shin*}
    \IEEEauthorblockA{
        \textit{School of Computer Science and Engineering} \\
        \textit{University of Aizu}, Fukushima, Japan \\
        jpshin@u-aizu.ac.jp
    }
    \and
    \IEEEauthorblockN{4\textsuperscript{th} Md. Al Mehedi Hasan}
    \IEEEauthorblockA{
        \textit{Department of Computer Science and Engineering} \\
        \textit{University of Engineering and Technology (RUET)} \\
        Rajshahi, Bangladesh \\
        mehedi\_ru@yahoo.com
    }
}

\maketitle
\begin{abstract}
Cervical cancer is a crucial global health concern for women, and  the persistent infection of High-risk HPV mainly triggers this remains a global health challenge, with young women’s diagnosis rates soaring from 10\% to 40\%
over three decades. While Pap smear screening is a prevalent diagnostic method, visual image analysis can be lengthy and often leads to mistakes. Early detection of the disease can contribute significantly to improving patient outcomes. In recent decades, many researchers have employed machine learning techniques that achieved promise in cervical cancer detection processes based on medical images. In recent years, many researchers have employed various deep-learning techniques to achieve high-performance accuracy in detecting cervical cancer but are still facing various challenges. This research proposes an innovative and novel
approach to automate cervical cancer image classification using Multi-Head Self-Attention (MHSA) and convolutional neural networks (CNNs). The proposed method leverages the strengths of both MHSA mechanisms and CNN to effectively capture both local and global features within cervical images in two streams. MHSA facilitates the model's ability to focus on relevant regions of interest, while CNN extracts hierarchical features that contribute to accurate classification. Finally, we combined the two stream features and fed them into the classification module to refine the feature and the classification. To evaluate the performance of the proposed approach, we used the SIPaKMeD dataset, which classifies cervical cells into five categories. Our model achieved a remarkable accuracy of 98.522\%. This performance has high recognition accuracy of medical image classification and holds promise for its applicability in other medical image recognition
tasks.
\end{abstract}

\section{Introduction}
Cervical cancer, primarily caused by the ongoing infection of the Human Papilloma Virus (HPV), is a significant health issue for women worldwide \cite{zhang2020cervical}. In 2012, there were 83,000 cases in developed countries, making it the 11th most common cancer, and 445,000 cases in developing countries, ranking it the second most common. The diagnosis rate of cervical cancer in younger women has increased from 10\% to 40\% over the last three decades \cite{zhang2020cervical}\cite{ferlay2015cancer}. The World Health Organization (WHO) has set a goal to reduce new cases of cervical cancer by more than 40\% and to decrease the number of related deaths by 5 million by 2050. On November 17, 2020, the WHO launched its Global Strategy to accelerate the eradication of cervical cancer, focusing on three key actions: vaccination, screening, and treatment \cite{PAHO}. Pap smear screening is a commonly used method for cervical cancer screening, where a speculum is inserted to open the vagina, and cells are collected from the cervix for laboratory analysis \cite{Pap}. However, visual image analysis is time-consuming, requires significant effort, and skilled personnel, making it inefficient and prone to errors. Therefore, automating this analysis is essential, and deep learning is being utilized to automate it in modern medicine \cite{alsubai2023privacy}. Deep learning enables efficient image recognition without the need for cell segmentation. Recent advances in computer vision and deep learning have facilitated the automation of medical image classification with improved accuracy levels \cite{9836398,rahim2020hand,zhang2017deeppap,egawa2023dynamic}.

CNNs have demonstrated high accuracy in image recognition tasks, including cervical cancer image classification \cite{alsubai2023privacy} \cite{sun2020automatically,miah2022bensignnet,shin2023rotation,shin2023korean,miah2023multi,miah2023multistage}. Attention-based networks have shown excellence in various domains, such as sign language and image classification \cite{shin2023korean,miah2023multi}. One of the biggest challenges in detecting cervical cancer through Pap smear analysis is achieving high accuracy and computational efficiency. Moreover, cervical morphology cells provide important information such as size, color, and shape. To address the challenges of accuracy and computational complexity, we introduce a novel model that combines a Multi-Head Self-Attention (MHSA) based transformer with a CNN framework, which is a modified version of the model described in \cite{shin2023korean}.

The contributions of our proposed research to automatic cervical cancer detection include:

\begin{itemize}
\item Proposing a combination of CNN and MHSA feature-based cervical cancer detection systems.
\item Utilizing several statistical techniques to preprocess the input dataset for generalization and high performance.
\item Implementing a grain module in the model for vision patch recognition and feeding the patch recognition information into two streams: one composed of the MHSAT transfer and the other composed of several layers of the CNN module.
\item Concatenating the two feature streams and feeding them into a classification module designed with an excellent combination of different deep learning layers. After evaluating the proposed model on the SIPaKMeD dataset, our model achieved a 98.522\% accuracy, demonstrating its superiority.
\end{itemize}

The paper is organized as follows: Section \ref{sec2} summarizes related work on cervical cancer image classification, with the dataset described in Section \ref{dataset}. Section \ref{sec3} elaborates on the data used, its preprocessing, and the architecture. Section \ref{sec4} analyzes the results of the experiments. Section \ref{sec5} concludes our research and discusses future work.

\section{Related Work} \label{sec2}
Several researchers have explored the use of machine learning algorithms\cite{miah2020motor,joy2020multiclass,9230773,kabir2023investigating} and deep learning techniques\cite{miah2023dynamic,miah2023multi,hossain2023stochastic} for the classification of cervical cancer images. Approaches based on Convolutional Neural Networks (CNN) have shown remarkable accuracy. For instance, \cite{sholik2023classification} proposed a method involving feature extraction using a CNN model, followed by feature reduction through Linear Discriminant Analysis (LDA). The extracted features were then used to train a k-Nearest Neighbors (k-NN) classifier, resulting in accurate predictions. This approach achieved an accuracy of 97.54\% on the SIPaKMeD dataset. Another study by \cite{alsubai2023privacy} utilized a simple 4-layer CNN model, obtaining an accuracy of 91.13\% on the same dataset. The simplicity of this model led to reduced computational costs and faster response times. Additionally, \cite{haraz2023deep} introduced a method involving feature extraction and preprocessing steps, achieving an accuracy of 96.8\% using Support Vector Machines (SVMs) on the SIPaKMeD dataset. Other classifiers also performed well, with a neural network (NN) achieving an accuracy of 95.8\%, and k-NN achieving 94.1\% accuracy. \cite{attallah2023cercan} proposed CerCan-Net, a framework integrating three CNNs with fewer parameters and layers compared to models like MobileNet, DarkNet-19, and ResNet-18, resulting in an accuracy of 97.7\% on the SIPaKMeD dataset. \cite{pramanik2023msenet} introduced the Mean and Standard Deviation-based Ensemble Network (MSENet), combining base classifiers Xception, Inception V3, and VGG-16. Using a 5-fold cross-validation scheme, MSENet achieved an accuracy of 97.21\% on the SIPaKMeD dataset. These studies highlight the diverse approaches in utilizing machine learning and deep learning techniques for cervical cancer image classification, contributing to improved diagnostics and patient care.

\subsection{Data Selection} \label{dataset} 
\begin{figure}[h]
    \centering
    \includegraphics[scale=0.9]{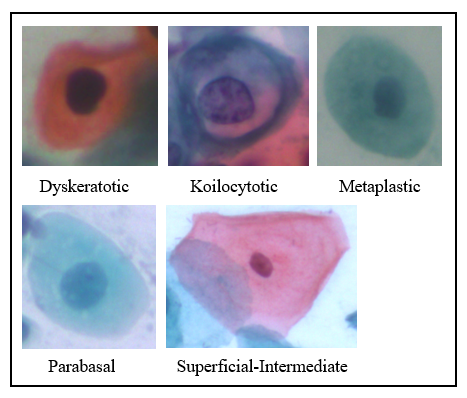}
    \caption{{SIPaKMeD Dataset Five Categories Images.}}
    \label{fig:Figure 1}
\end{figure}

The Pap smear is a primary method for cervical cancer screening. For this study, we use the SIPaKMeD dataset, a labeled, publicly available, and widely used dataset for cervical cancer image classification. The SIPaKMeD dataset was chosen for its comprehensiveness, as most alternative datasets lack sufficient volume for our model to demonstrate its full efficacy. This dataset contains 4049 cell images extracted from 966 cluster cell images of Pap smear slides, classified into five categories (Fig. \ref{fig:Figure 1}): Superficial-Intermediate, Parabasal, Koilocytotic, Dyskeratotic, and Metaplastic \cite{plissiti2018sipakmed}. The images were resized to a 192 × 192 pixel size. Each cell type has distinct morphological characteristics, providing valuable information for classification and diagnosis.

\section{Proposed Methodology} \label{sec3}
\begin{figure*}[t] 
    \centering
    \includegraphics[scale=0.40]{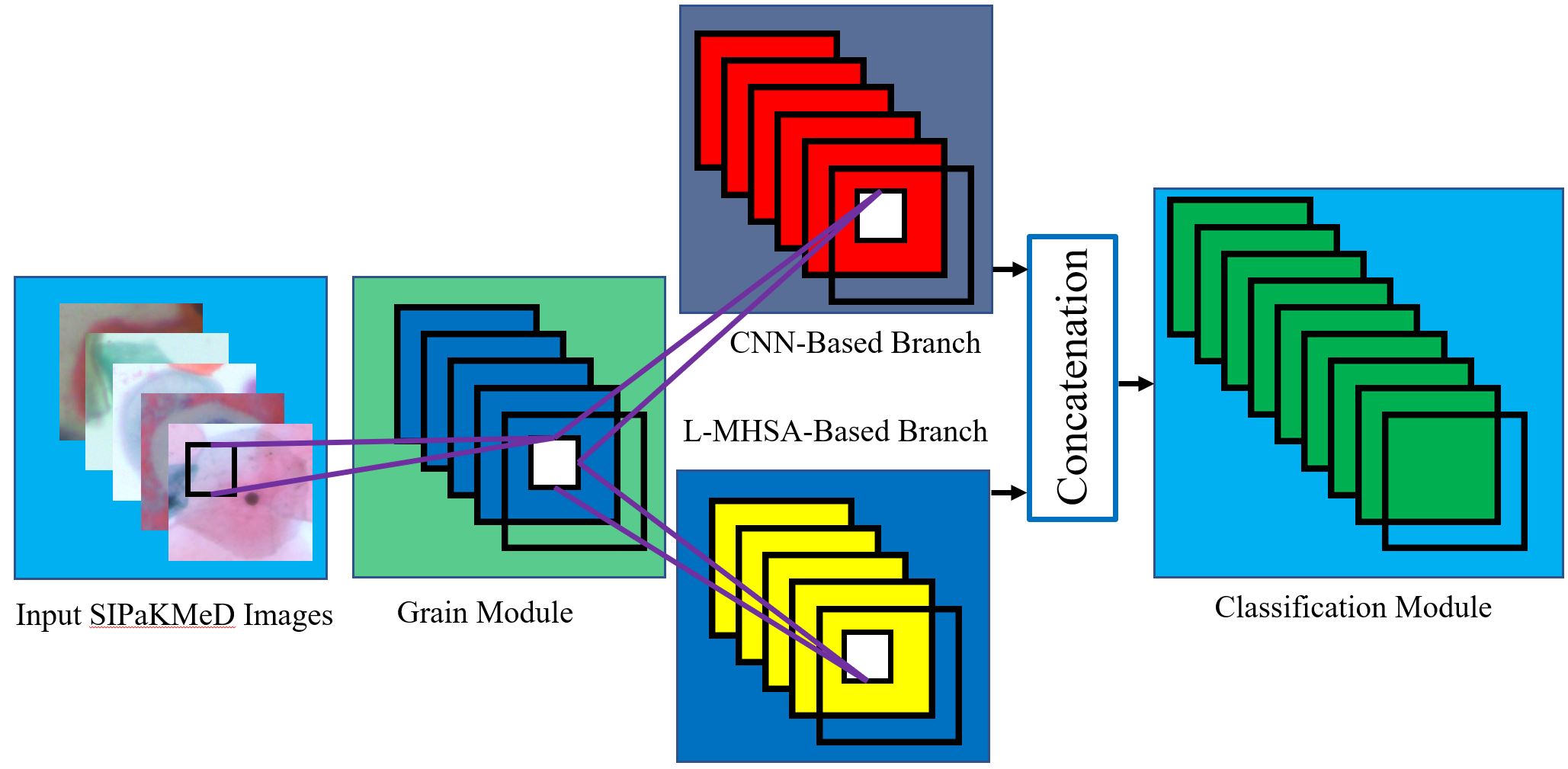}
    \caption{{Our Proposed Model Architecture.}}
    \label{fig:Figure 2} 
\end{figure*}

In our study, we introduced a multi-stream model combining the Multi-Head Self-Attention (MHSA) and Convolutional Neural Network (CNN) modules, an advanced version of the system presented in \cite{shin2023korean}, which achieved high accuracy in the sign language recognition task \cite{shin2024korean_ksl0,10360810_miah_ksl2,miah2024hand_multiculture,miah2024sign_largescale,miah2024spatial_paa,miah2023multi}. The motivation behind this approach is the limited use of attention-based systems in the recognition of cervical cancer images. The working flow diagram of the proposed model is depicted in Fig. \ref{fig:Figure 2}. The model consists of a multi-branch hybrid network that integrates the MHSA transform and CNN, drawing upon the architectures of CMT \cite{guo2022cmt}, ResNet-50 \cite{he2016deep}, and DeiT \cite{touvron2021training}\cite{cui2019deep}. This combination aims to leverage the strengths of both attention mechanisms and convolutional layers to improve the accuracy of cervical cancer image classification.


\subsubsection{Grain Module}
In the first stage of the proposed model, a grain module is employed to generate fine-grained features of the dataset. These fine-grained features are then fed into two streams, namely the MHSA and CNN modules. The input image is directed into the grain module to extract delicate initial features. This module is divided into three segments. The first segment utilizes a 3 × 3 convolution with a stride of 2, adjusting the output channels to 32 and thereby reducing the size of the input image. Subsequently, two additional 3 × 3 convolution layers with a stride of 1 are applied. In the second segment, a patch aggregation approach is carried out using convolutional layers and layer normalization.

\begin{figure}[h]
    \centering
    \includegraphics[scale=0.25]{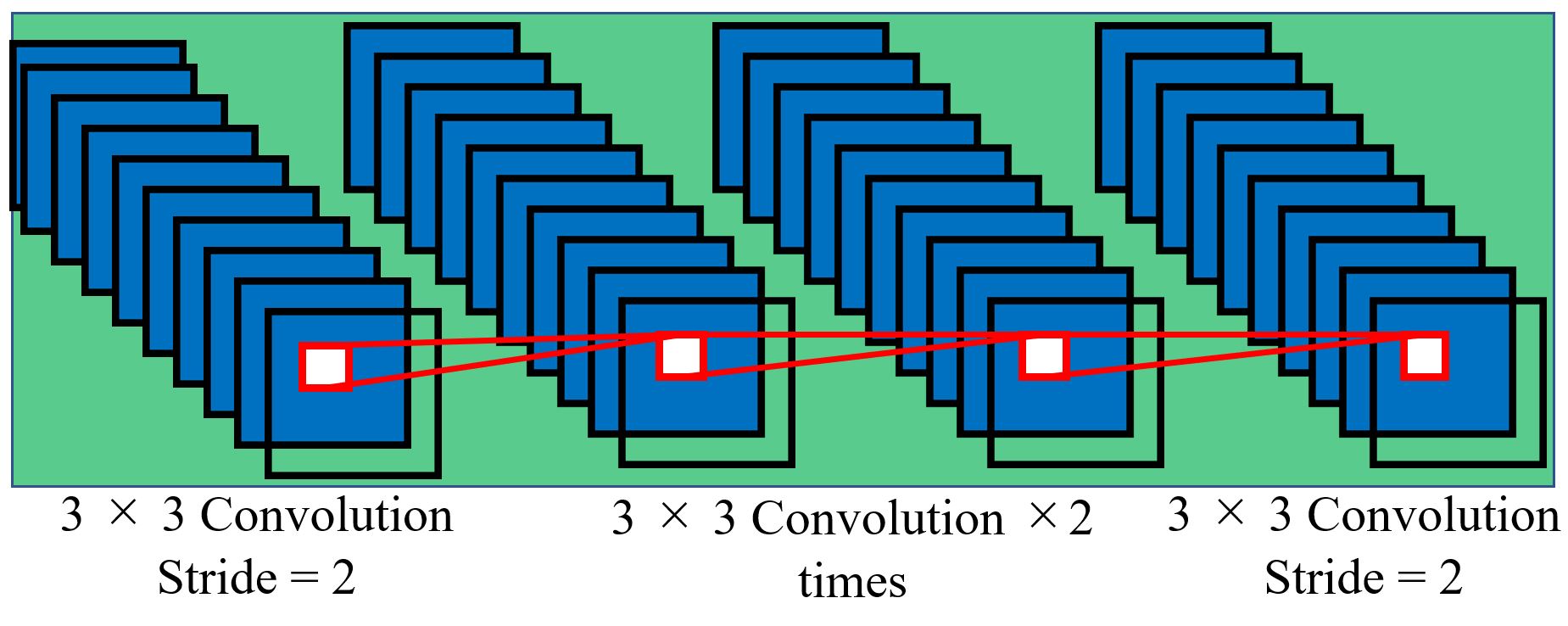}
    \caption{{The Grain Module Architecture.}}
    \label{fig:Figure 3}
\end{figure}

The architecture of the Grain Module is shown in Fig. \ref{fig:Figure 3}.
\subsubsection{MHSA based Stream}
The grain module in our proposed model is structured into three key segments: an initiating segment, a Lightweight Multi-Head Self-Attention (LMHSA) segment \cite{guo2022cmt}, and MLP convolution. The initiating segment is designed to extract localized information from the dataset, incorporating position encoding to capture spatial details. A primary objective of this segment is to accommodate augmentation practices such as rotation and shift, which are crucial for visual operations. By integrating these practices, the module aims to mitigate the dependency on image transformations, ensuring that the other components of the system remain unaffected by such variations.

\begin{figure*}[t] 
    \centering
    \includegraphics[scale=0.25]{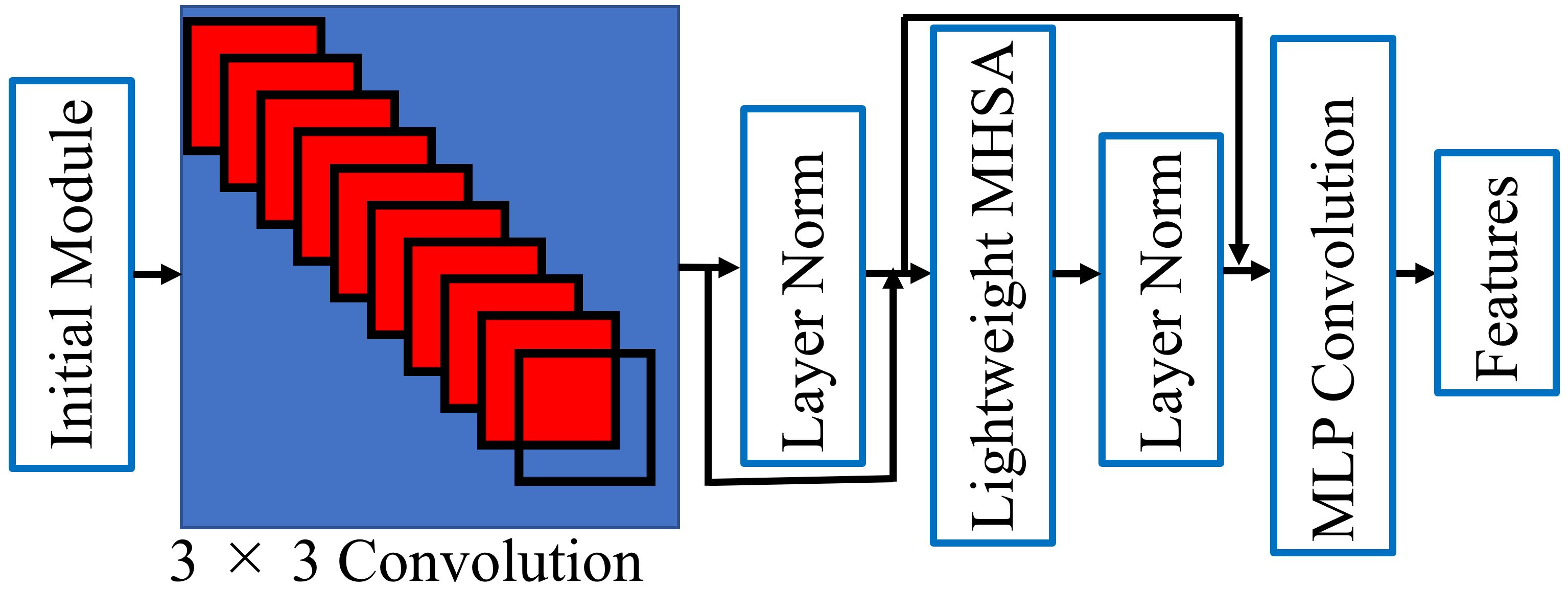}
    \caption{{The Convolutional Layer-Based Transformer Module Architecture.}}
    \label{fig:Figure 4} 
\end{figure*}

The architecture of the Grain Module is shown in Fig. \ref{fig:Figure 4}.

\paragraph{Initialization Module}
The primary function of this stream is to employ augmentation techniques such as rotation, shifting, and translational dependency. While many studies have used various kinds of augmentation techniques, including manual mathematical ranges, our study employs a deep learning-based augmentation technique. This technique is designed to address issues related to shifting, rotation, and translation dependency, thereby enhancing the generalization property of the model. It emphasizes crucial visual operations, such as rotation and shift enhancements, ensuring that these operations do not disrupt the overall system. Conventional transformers often use absolute positional encoding, which can overlook local interconnections and intra-patch configurations. Our proposed approach aims to rectify these oversights, providing a more effective and comprehensive solution for image analysis.

\paragraph{LMHSA Module}
The output from the initialization module is directed into the LMHSA module, which generates four distinct patterns in our case. The concept of self-attention (SA) stands out as a pivotal element within neural network architectures. It functions by segmenting input data into three separate matrices: query, key, and value. Our approach seeks to boost the efficiency of this process by incorporating an element-wise convolutional neural network. This integration fulfills a dual purpose: it assists in reducing the dimensions for both the key and value components, and it facilitates a more streamlined attention step.
A notable enhancement in our approach is the addition of a relative positional bias within each self-attention module. This innovative feature leads to the development of what we refer to as the Lightweight Self Attention (LSA) design. The LSA design is a blend of structural elegance and advanced processing, poised to potentially transform the landscape of self-attention mechanisms by introducing a spatial awareness element.
The implementation and optimization of the LSA model were carried out with meticulous attention to detail. The combination of element-wise convolutional neural networks and the strategic inclusion of positional bias were executed with precision. The result is a neural architecture that not only demonstrates efficiency in terms of computational resources but also exhibits an enhanced ability to capture complex relationships within the data.
Furthermore, the LSA design proves its effectiveness across a wide range of applications. From natural language processing tasks that demand an understanding of contextual subtleties to computer vision challenges that require the extraction of intricate features, the LSA model showcases its versatility. This adaptability highlights the potential of self-attention mechanisms to break through traditional boundaries and establish a presence in various domains.
In conclusion, the integration of self-attention (SA) within neural networks has paved the way for improved processing capabilities. Through the combination of convolutional neural networks and the incorporation of relative positional bias, our Lightweight Self Attention (LSA) design has emerged. This design encapsulates both efficiency and sophistication, offering a glimpse into the future trajectory of self-attention mechanisms. As research and innovation continue to advance, the impact of such developments is felt across multiple fields, promising innovative solutions and higher performance standards.

\paragraph{MLP Convolution}
After the attention process in the MHSA, our architecture seamlessly integrates an MLP convolution block, consisting of two distinct 1 × 1 convolution layers. This integration marks a significant departure from traditional methods, endowing our model with a unique capability. Unlike conventional convolution layers that primarily focus on local patterns, the MLP convolution layer enhances our model's ability to concentrate on specific positions within the data.
A key aspect that distinguishes the MLP convolution from its conventional 2D counterpart is the orientation of channels. While 2D convolution layers are adept at extracting spatial structures and relationships, MLP convolution layers serve a different purpose. They act as translators, transforming the broader-scale data representations produced by the MHSA into finely-grained pixel-level information. This distinction lies at the heart of our approach, illustrating the intricate interplay between these two convolution methodologies.
The combination of MHSA and MLP convolution forms a dynamic synergy, effectively merging the global context-capturing capabilities of MHSA with the pixel-level precision of MLP convolution. This synergy highlights our architecture's proficiency in managing complex data patterns and hierarchies.
It is important to emphasize the meticulous calibration and fine-tuning that were essential in harmonizing these components. The seamless integration of MHSA and MLP convolution involved exploring optimal hyperparameters, architectural nuances, and validation strategies. The result of this careful endeavor is a neural model capable of unraveling intricate data complexities while demonstrating commendable efficiency.
Furthermore, the implications of this hybrid approach extend beyond a single application domain. The versatility of our architecture is evident across various fields. In natural language processing, the ability to translate broad semantic context into nuanced word-level details could revolutionize sentiment analysis, machine translation, and contextual understanding. In computer vision, the combination of global feature extraction and pixel-level precision holds promise for object recognition, image segmentation, and generative tasks.
In conclusion, the incorporation of an MLP convolution block following the attention process in MHSA represents a novel advancement in neural network architecture. This unique approach leverages the channel orientation of MLP convolution to convert MHSA's global context into detailed pixel-level insights. The resulting synergy combines the strengths of both methodologies, positioning our model as a powerful tool for tackling complex data-driven challenges across various domains. As the research landscape continues to evolve, the impact of such innovative combinations resonates, shaping the future of artificial intelligence applications.

\subsubsection{CNN Module}
To enhance the local feature from the previous MLP convolution module, we used a CNN module. It is used to draw out local characteristics from the grain feature; we utilized four sequences of 3 × 3 convolution layers paired with GELU activation and batch normalization. This CNN module design can have a notable impact on the performance of a neural network, particularly in tasks such as image classification, object detection, and segmentation. The sequence of convolution layers, activation functions, and batch normalization introduces a series of transformations to the input data. The use of 3 × 3 convolutional filters enables the network to capture local patterns and features effectively. The GELU activation function, known for its smoothness and effectiveness in deep networks, contributes to the network's ability to model complex non-linear relationships within the data. Batch normalization aids in stabilizing and accelerating training by normalizing the activation and reducing internal covariate shift, and the reason for the use of the CNN layer is described below.
\begin{itemize}
\item Feature Extraction: The sequence of 3 × 3 convolutional layers allows the module to extract hierarchical features from the input data progressively. Each layer captures different levels of abstraction, enabling the network to learn intricate patterns and structures.
\item GELU Activation: The GELU activation function helps mitigate the vanishing gradient problem and accelerates convergence during training. Its smoothness promotes more efficient optimization and can contribute to better generalization.
\item Batch Normalization: The incorporation of batch normalization reduces internal covariate shift, leading to faster convergence during training. It also acts as a regularizer, contributing to improved model generalization and reduced overfitting.
\item Parameter Efficiency: The use of 3 × 3 convolutional filters reduces the number of parameters compared to larger filters, enabling the model to capture a wide range of patterns with relatively fewer parameters.
\end{itemize}

\begin{figure}[h]
    \centering
    \includegraphics[scale=0.20]{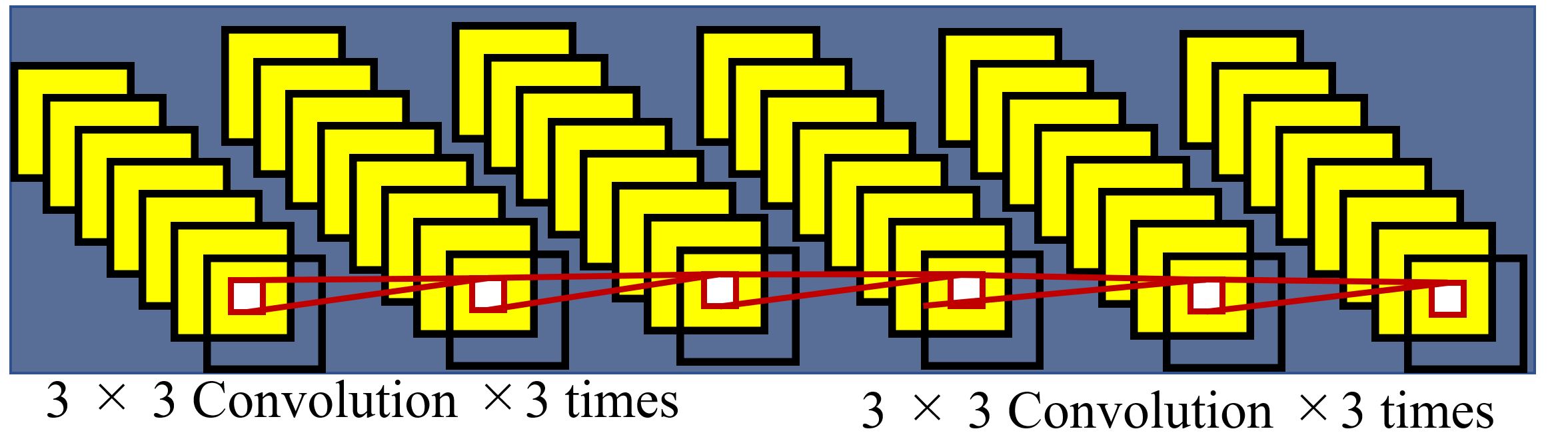}
    \caption{{The CNN Module Architecture.}}
    \label{fig:Figure 5}
\end{figure}

The architecture of the CNN Module is shown in Fig. \ref{fig:Figure 5}.

\subsubsection{Classification Module}
In the classification module of our architecture, we employed a modified version of the Feed-Forward Network (FFN) from the Vision Transformer (ViT) \cite{loshchilov2017decoupled}. This modified FFN is characterized by two linear layers, which are separated by the GELU \cite{de2020sign} activation function. The first linear layer expands the dimensionality of the input by a factor of four, while the second linear layer reduces it back to the original size. This design is similar to the Inverted Residual Feed-Forward Network (IRFFN) \cite{guo2022cmt}\cite{chen2021pre}, which includes an expansion layer implemented through element-wise convolution, followed by a pair of projection layers that reshape the dimensions. The use of this tweaked FFN in the classification module is instrumental in enhancing the model's ability to capture and process complex patterns within the data, ultimately contributing to improved classification performance.

\begin{figure}[h]
    \centering
    \includegraphics[scale=0.20]{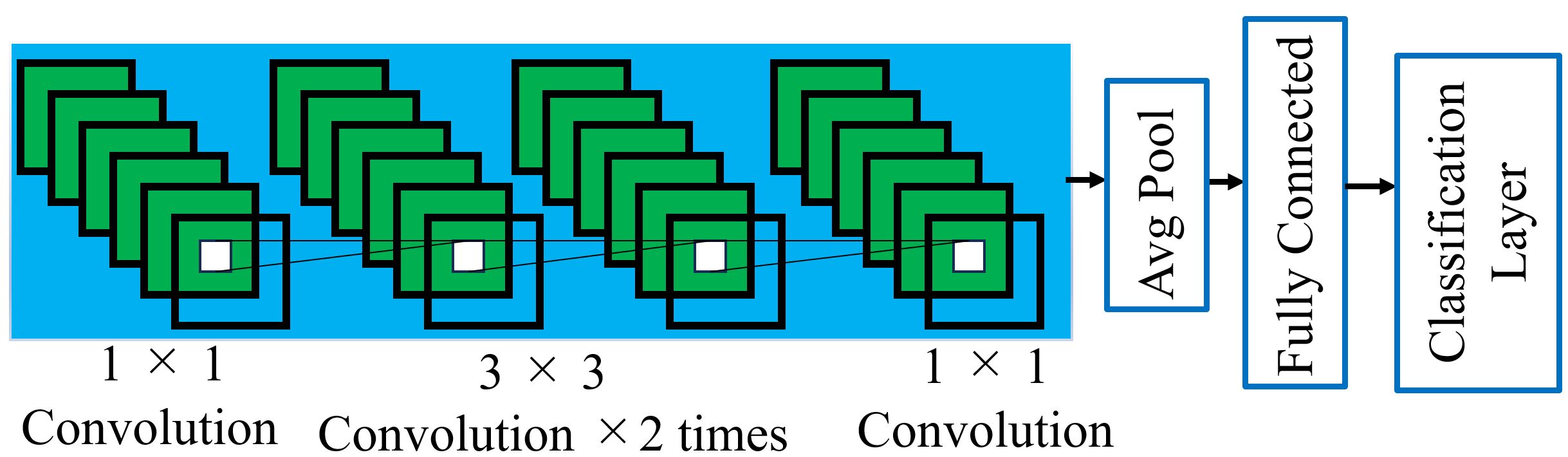}
    \caption{{The Classification Module Architecture.}}
    \label{fig:Figure 6}
\end{figure}

The architecture of the Grain Module is shown in Fig. \ref{fig:Figure 6}.

The overall goal of the proposed architecture is to harness the strengths of both transformers and CNNs, effectively combining their capabilities for enhanced performance in tasks like image analysis.

\section{Experimental Evaluation} \label{sec4} 
We evaluated our model using the SIPaKMeD dataset, a benchmark cervical cancer dataset mentioned earlier. This dataset was chosen to test the effectiveness of the proposed model. We trained the model on 80\% of the dataset and used the remaining 20\% for testing. During the training phase, specific parameters were adjusted to optimize the learning process. These parameters included a warmup epoch of 5, a weight decay of 0.00000001, a base learning rate of 0.001, a warmup learning rate of 0.00000002, and a minimum learning rate of 0.0002. These settings were based on empirical evidence and previous experiences with image classification. A batch size of 64 was used to enable efficient parallel processing. The training environment utilized CUDA version 11.6, and the model was trained for a total of 2000 epochs using the AdamW optimizer.

\subsection{Performance Result}
Through our thorough training regimen, we observed a significant increase in test accuracy. The accuracy improved from 61.084\% initially to an impressive 98.522\%. This progress showed a clear upward trend, reaching a plateau at the 1539th epoch, indicating that further training beyond this point was unnecessary. The evolution of accuracy over each epoch is depicted in the associated figure. This meticulous approach to model training, parameter tuning, and accuracy analysis highlights the careful planning behind this experiment. The substantial improvement in accuracy not only demonstrates the effectiveness of the proposed model but also highlights the potential for enhancing precision through systematic training. This success underscores the importance of advanced computational techniques in addressing critical healthcare challenges.

\begin{figure}[h]
    \centering
    \includegraphics[scale=0.5]{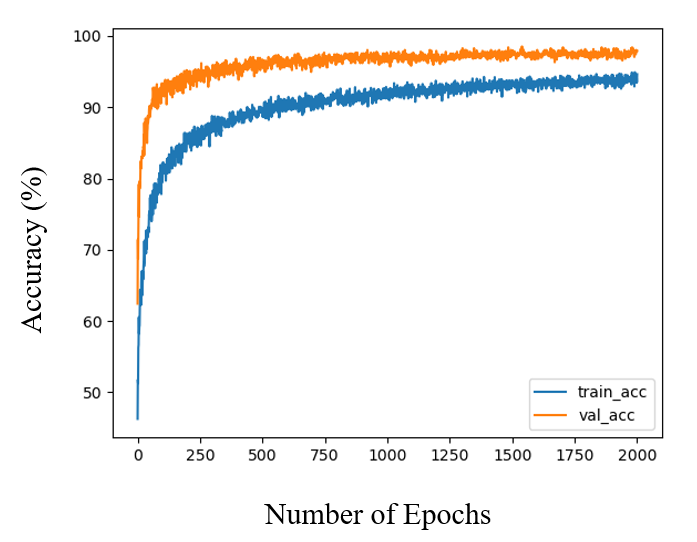}
    \caption{{The accuracy of train and test per epoch.}}
    \label{fig:Figure 7}
\end{figure}

The accuracy for each epoch is shown in Fig.  \ref{fig:Figure 7}.
\subsection{State of the Art Comparison of the Proposed Model}
Table \ref{tab:Table 1} presents a comparison of the accuracy achieved by our method with that of other research. The simple SVM method \cite{haraz2023deep} reached an accuracy of 96.8\%. The CerCan-Net method \cite{attallah2023cercan}, which integrates features from three CNN layers, achieved 97.7\%. The MSENet method \cite{pramanik2023msenet}, which combines a model consisting of three fundamental base classifiers: Xception, Inception V3, and VGG-16, achieved 97.21\%. Our method surpassed the highest accuracy among these by 0.822\%, demonstrating its effectiveness in cervical cancer image classification.

\begin{table}[h]
    \centering
    \caption{The accuracy comparison with other research.}
    \label{tab:accuracy_comparison}
    \begin{tabular}{|l|l|l|}
    \hline
        Method & Accuracy (\%) & Dataset \\
        \cline{1-3} \\[-1ex]
        CNN + LDA + k-NN\cite{sholik2023classification} & 97.54 & SIPaKMeD \\
        4-layer CNN\cite{alsubai2023privacy} & 91.13 & SIPaKMeD \\
        SVM\cite{haraz2023deep} & 96.8 & SIPaKMeD \\
        CerCan-Net\cite{attallah2023cercan} & 97.7 & SIPaKMeD \\
        MSENet\cite{pramanik2023msenet} & 97.21 & SIPaKMeD \\
        Our Method & 98.52 & SIPaKMeD \\ \hline
    \end{tabular}
    \label{tab:Table 1}
\end{table}
\section{Conclusions} \label{sec5}
Cervical cancer screening, which involves the manual classification of cell images, is inefficient and requires a significant investment of time and personnel. With the rising global incidence of cervical cancer, the need to automate and streamline the screening process is clear. Our study provides strong evidence of the effectiveness of our proposed model in classifying cervical cancer images, achieving an impressive accuracy rate of 98.522\% on the SIPaKMeD dataset. The exceptional performance of our model in two distinct tasks, sign language recognition and cervical cancer image classification, highlights its versatile potential. This success opens the door to applying the model to various other medical image recognition tasks. By refining the model's architecture and training it on extensive datasets, the possibility of achieving high accuracy in different medical contexts becomes increasingly feasible. This holds the potential to revolutionize medical diagnostics and demonstrates the synergy between advancements in artificial intelligence and the imperative to improve healthcare outcomes.


\vspace{12pt}

\bibliography{ref}
\bibliographystyle{plain}

\end{document}